\documentclass[pra,twocolumn,showpacs]{revtex4}
\usepackage{graphicx}

\newcommand{\beq}{\begin{equation}}
\newcommand{\eeq}{\end{equation}}
\newcommand{\beqa}{\begin{eqnarray}}
\newcommand{\eeqa}{\end{eqnarray}}
\newcommand{\ba}{\begin{array}}
\newcommand{\ea}{\end{array}}

\begin{document}

\title{Matter-wave vortices in cigar-shaped and toroidal waveguides}
\author{L. Salasnich$^{1}$, B. A. Malomed$^{2}$ and F. Toigo$^{1}$}
\affiliation{$^1$CNISM and CNR-INFM, Unit\`a di Padova, \\
Dipartimento di Fisica ``Galileo Galilei'', Universit\`a di Padova, Via
Marzolo 8, 35131 Padova, Italy \\
$^{2}$Department of Physical Electronics, School of Electrical Engineering,
Faculty of Engineering, Tel Aviv University, Tel Aviv 69978, Israel}

\begin{abstract}
We study vortical states in a Bose-Einstein condensate (BEC) filling a
cigar-shaped trap. An effective one-dimensional (1D) nonpolynomial Schr\"{o}%
dinger equation (NPSE) is derived in this setting, for the models with both
repulsive and attractive inter-atomic interactions. Analytical formulas for
the density profiles are obtained from the NPSE in the case of
self-repulsion within the Thomas-Fermi approximation, and in the case of the
self-attraction as exact solutions (bright solitons). A crucially important
ingredient of the analysis is the comparison of these predictions with direct
numerical solutions for the vortex states in the underlying 3D
Gross-Pitaevskii equation (GPE). The comparison demonstrates that the NPSE
provides for a very accurate approximation, in all the cases, including the
prediction of the stability of the bright solitons and collapse threshold
for them. In addition to the straight cigar-shaped trap, we also consider a
torus-shaped configuration. In that case, we find a threshold for the
transition from the axially uniform state, with the transverse intrinsic
vorticity, to a symmetry-breaking pattern, due to the instability in the
self-attractive BEC filling the circular trap.
\end{abstract}

\maketitle

\section{Introduction}

An important setting for experimental and theoretical studies of dynamical
phenomena in Bose-Einstein condensates (BECs) is provided by nearly
one-dimensional (1D) \textquotedblleft cigar-shaped" traps, which assumes
tight confinement in the transverse plane, allowing to unravel the dynamics
along the longitudinal axis. The use of this geometry helped to achieve
famous experimental results, such as the creation of single \cite%
{Khaykovich02} and multiple \cite{Strecker02} bright solitons in the
condensate of $\ $several thousand $^{7}$Li atoms. In those experiments, the
strength of the interaction between atoms was controlled and made weakly
attractive by means of the Feshbach-resonance technique. In the condensate
of $^{85}$Rb atoms trapped in a similar geometry, nearly 3D solitons were
found in a post-collapse state \cite{Cornish}.

It is natural to derive effectively one-dimensional (1D) equation(s) for the
description of this experimentally relevant setting, starting from the full
3D Gross-Pitaevskii equation (GPE). The reduction of the 3D equation to a 1D
form was performed, under various assumptions, by means of sundry methods 
\cite{PerezGarcia98}-\cite{boris}. In the simplest approximation, the
difference of the eventual equation from a formal 1D variant of the GPE
(with the underlying cubic nonlinearity) is represented by an additional
quintic term, whose sign is always self-attractive, irrespective of the sign
of the cubic nonlinearity \cite{Shlyap02}. A more consistent approach to the
derivation of the 1D equation postulates the factorization of the 3D
mean-field wave function into the product of the ground state of the 2D
harmonic oscillator in the transverse plane, and slowly varying axial (1D)
wave function. Using the variational representation of the underlying 3D
GPE, this approach ends up with the \textit{nonpolynomial Schr\"{o}dinger
equation} (NPSE) for the axial wave function \cite{sala1,sala2,sala3,sala5}.
The NPSE has been extended to investigate the Tonks-Girardeau regime \cite%
{sala-tonks}, the two-component BEC \cite{sala4}, and also transverse
spatial modulations \cite{sala6}.

Another physically interesting modification of the above setting is that
when the BEC trapped in the cigar-shaped geometry is lent vorticity
(assuming that the corresponding vector of the angular momentum is directed
parallel to axis of the cigar-shaped trap), as proposed in Ref. \cite%
{sala-vortex}. In the experiment, the vorticity may be imparted on the
condensate by a helical laser beam shining along the axis of the trap. A
natural problem for the theoretical analysis, which is considered in the
present work, is the derivation of a modification of the effective
one-dimensional NPSE that takes into regard the intrinsic vorticity. As we
demonstrate in Section II, such a modified 1D equation is not drastically
different from its counterpart derived before \cite{sala1,sala2} for the
zero-vorticity states. However, a really nontrivial issue is to compare
basic dynamical states predicted by this effective equation with direct
numerical solutions of the underlying GPE in three dimensions. In this work,
we perform the comparison separately for the self-repulsive and
self-attractive BEC (in Sections III and IV, respectively). In the former
case, the relevant states are of the Thomas-Fermi (TF) type, while in the
latter model (self-attraction) these are bright solitons. We demonstrate
that the modified NPSE provides for very good approximation for all these
states, including the prediction of their stability (and, in particular, of
the collapse threshold for solitons in the self-attraction model). In
addition, in Section V we consider a different geometry, when the long
quasi-1D trap is bent and closed into a ring (torus), maintaining the
intrinsic vorticity in the transverse plane. In that case, we find that the
modified NPSE accurately predicts the shape and stability of a possible
soliton, as well as the delocalization transition to the state of the BEC
uniformly filling the toroidal trap.

\section{The nonpolynomial Schr\"{o}dinger equation (NPSE) for matter-wave
vortices}

The description of a dilute BEC, confined in the axial direction by a
generic potential $V(z)$, and in the transverse plane by the harmonic
potential with frequency $\omega _{\bot }$, is based on the fundamental GPE
in three dimensions 
\begin{equation}
i{\frac{\partial \psi }{\partial t}}=\left[ -{\frac{1}{2}}\nabla^{2} +{\frac{%
1}{2}}(x^{2}+y^{2})+V(z)+2\pi g|\psi |^{2}\right] \psi \;,  \label{3DGPE}
\end{equation}
where $\psi (\mathbf{r},t)$ is the macroscopic wave function of the
condensate normalized to unity, and $g\equiv 2Na/a_{\bot }$, with $N$ the
number of atoms, $a_{\bot }=\sqrt{\hbar /(m\omega _{\bot })}$ the length of
the transverse harmonic confinement, and $a$ the scattering length of atomic
collisions. In Eq. (\ref{3DGPE}), the length and time units are $a_{\bot }$
and $\omega _{\bot }^{-1}$, and the energy unit is $\hbar \omega _{\bot }$.
This equation can be derived from the Lagrangian density:  
\begin{eqnarray}
\mathcal{L} &=&\frac{i}{2}\left( \psi ^{\ast }{\frac{\partial \psi }{%
\partial t}-}\psi {\frac{\partial \psi ^{\ast }}{\partial t}}\right) -{\frac{%
1}{2}}|\nabla \psi |^{2}-{\frac{1}{2}}(x^{2}+y^{2})|\psi |^{2}  \nonumber \\
&&-V(z)|\psi |^{2}-\pi g|\psi |^{4}\;.  \label{lagrangian}
\end{eqnarray}

In this work, we study the existence and properties of matter-wave states
with vorticity in the transverse plane, which correspond to solutions of the
form 
\begin{equation}
\psi (\mathbf{r},t)=\Phi (r,z,t)\ e^{iS\theta }\;,  \label{vortex}
\end{equation}%
where $S$ is the integer vorticity quantum number, $r$ and $\theta $ being
the polar coordinates in the $\left( x,y\right) $ plane. On the substitution
of expression (\ref{vortex}), Eq. (\ref{3DGPE}) takes the form: 
\begin{eqnarray}
i{\frac{\partial \Phi }{\partial t}} &=&\left[ -{\frac{1}{2}}\left( {\frac{%
\partial ^{2}}{\partial r^{2}}}+{\frac{1}{r}}{\frac{\partial }{\partial r}}+{%
\frac{\partial ^{2}}{\partial z^{2}}}\right) \right.   \nonumber \\
&&\left. +{\frac{S^{2}}{2r^{2}}}+{\frac{1}{2}}r^{2}+V(z)+2\pi g|\Phi |^{2}%
\right] \Phi \;.  \label{3DGPE-vortex}
\end{eqnarray}%
We solved Eq. (\ref{3DGPE-vortex}) numerically by using the
finite-difference Crank-Nicholson predictor-corrector method with the
cylindric symmetry described in Ref. \cite{sala-numerics}. The 
Crank-Nicholson algorithm with imaginary time is used to find the
stationary wave function of the GPE with a fixed vorticity $S$. 
The algorithm provides for fast convergence to a finite-size 
wave function with finite energy. In the case of instability 
due to collapse ($g<0$) the algorithm quickly approaches a 
delta peak with (negative) diverging energy.
Note that, in the free space, vortices are usually destroyed by the
instability against azimuthal perturbations \cite{DumDum},
which split them into a set of separating fragments \cite{newref},
but in our setting this instability is suppressed
by the tight transverse confinement. 

To gain some analytical insight and to reduce the problem to 1-D, we apply a
variational approach. In the case of a cigar-shaped geometry, it is natural
to extend the variational representation \cite{sala1,sala2} of the 3D GPE
which led to the NPSE for the axial wave function in the case of no
vorticity to the present situation. Thus we adopt the following ansatz for
the vortex state described by Eq.(\ref{vortex}): 
\begin{equation}
\Phi (r,z,t)={\frac{r^{S}}{\sqrt{\pi S!}\sigma (z,t)^{S+1}}} \exp{\left[ -{\ 
\frac{r^{2}}{2\sigma (z,t)^{2}}}\right] }\,f(z,t)\; ,  \label{ansatz-npse}
\end{equation}
where $\sigma (z,t)$ and $f(z,t)$ which account for the transverse width and
the amplitude of the vortex are to be determined variationally. 
By inserting this
ansatz into the Lagrangian density (\ref{lagrangian}) and performing the
integration over $x$ and $y$, we derive the effective Lagrangian density, 
\begin{eqnarray}
\bar{\mathcal{L}} &=&if^{\ast }{\frac{\partial f}{\partial t}}-{\frac{1}{2}}
\left\vert {\frac{\partial f}{\partial z}}\right\vert ^{2}-{\frac{1}{2}}
(S+1)\left( {\frac{1}{\sigma ^{2}}}+\sigma ^{2}\right) |f|^{2}  \nonumber \\
&& -{\frac{(S+1)}{2\sigma ^{2}}}\left( {\frac{\partial \sigma }{\partial z}}
\right) ^{2}|f|^{2}-V(z)|f|^{2}-{\frac{1}{2}}(S+1){g}_{S}{\frac{|f|^{4}}{
\sigma ^{2}}}\;,  \label{effective}
\end{eqnarray}
where the effective nonlinearity strength for the vortex state is 
\[
g_{S}=g\ {\frac{(2S)!}{2^{2S}(S+1)(S!)^{2}}}\;. 
\]%
Note that the effective Lagrangian density (\ref{effective}) does not
contain time derivatives of $\sigma (z,t)$. Expression (\ref{effective})
gives rise to the corresponding Euler-Lagrange equations,%
\begin{eqnarray}
i{\frac{\partial f}{\partial t}} &=&\left[ -{\frac{1}{2}}{\frac{\partial^{2} 
}{\partial z^{2}}}+V(z)+{\frac{1}{2}}(S+1)\left( {\frac{1}{\sigma ^{2}}}
+\sigma ^{2}\right) \right.  \nonumber \\
&&\left. +{\frac{(S+1)}{2\sigma ^{2}}}\sigma ^{\prime 2}+{\frac{
g_{S}(S+1)|f|^{2}}{\sigma ^{2}}}\right] f\;,  \label{eq1}
\end{eqnarray}
\begin{equation}
\sigma ^{4}=1+g_{S}|f|^{2}+\sigma ^{\prime 2}+{\frac{\sigma ^{3}}{|f|^{2}}} {%
\frac{\partial }{\partial z}}\left( {\frac{\sigma ^{\prime }}{\sigma ^{2}}}
|f|^{2}\right) \;,  \label{eq2}
\end{equation}
where $\sigma^{\prime }\equiv \partial \sigma /\partial z$. Neglecting this
derivative, Eqs. (\ref{eq1}) and (\ref{eq2}) can be combined into:  
\begin{equation}
i{\frac{\partial f}{\partial t}} = \left[ -{\frac{1}{2}}{\frac{\partial ^{2}%
} {\partial z^{2}}}+V(z)+(S+1){\frac{1+(3/2)g_{S} |f|^{2}}{\sqrt{%
1+g_{S}|f|^{2}}}}\right] f\;,  \label{npse}
\end{equation}%
which is a new version of the nonpolynomial Schr\"{o}dinger equation (NPSE)
for the vortical state tightly confined in the transverse plane. In the case
of $S=0$, Eq. (\ref{npse}) reduces to the ordinary NPSE, that has found
various applications \cite{sala1,sala6}.

The eigenfunctions of the equation 
\begin{equation}
\left[ -{\frac{1}{2}}{\frac{\partial ^{2}}{\partial z^{2}}}+V(z)+(S+1){\frac{%
1+(3/2)g_{S}|\phi |^{2}}{\sqrt{1+g_{S}|\phi |^{2}}}}\right] \phi =\mu \ \phi 
\label{npse-st}
\end{equation}%
provide the stationary solutions of Eq. (\ref{npse}) 
of the form $f(z,t)=\phi (z)\ e^{-i\mu t}$. Obviously, Eqs. (\ref%
{npse}) and (\ref{npse-st}) can be transformed into their counterparts for $%
S=0$ by substitutions $g\rightarrow g_{S}$, $\mu \rightarrow \mu /(S+1)$, $%
t\rightarrow \left( S+1\right) t$, $z\rightarrow \sqrt{S+1}z$, and $%
V\rightarrow V/\left( S+1\right) $. Therefore, the results obtained in
earlier works in the framework of the NPSE with $S=0$ can be easily applied
to Eqs. (\ref{npse}) and (\ref{npse-st}). A really nontrivial issue is to
compare these results with those produced by numerical integration of the
full three-dimensional GPE for vortical states.

\section{Repulsive nonlinearity}

In the case of a repulsive inter-atomic interactions 
(i.e. $g>0$) one may readly generalized the TF approximation 
developed in Ref. \cite{sala3}, for $S=0$ to the case with vorticity $S$. 
Thus, neglecting the kinetic-energy term, i.e., 
the second derivative, in Eqs. (\ref%
{npse-st}) one gets the analytical expression for the stationary axial
probability density $\rho (z)=|\phi (z)|^{2}$ in the form:  
\begin{equation}
\rho (z)={\frac{2}{9g_{S}}}\left[ \mu _{S}^{2}(z)-3+\mu _{S}(z)\sqrt{\mu
_{S}^{2}(z)+3}\right] ,  \label{figata}
\end{equation}%
where the effective \textit{local} chemical potential is 
\begin{equation}
\mu_{S}(z)={\frac{\mu -V(z)}{S+1}}  \label{figata1} \; . 
\end{equation}%
Notice that generally expression (\ref{figata}) does not represent a
soliton. Equations (\ref{figata}) and (\ref{figata1}) are a straightforward
generalization of the TF approximation developed in Ref. \cite{sala3} for
the repulsive BEC with $S=0$.

\begin{figure}[tbp]
{\includegraphics[height=2.3in,clip]{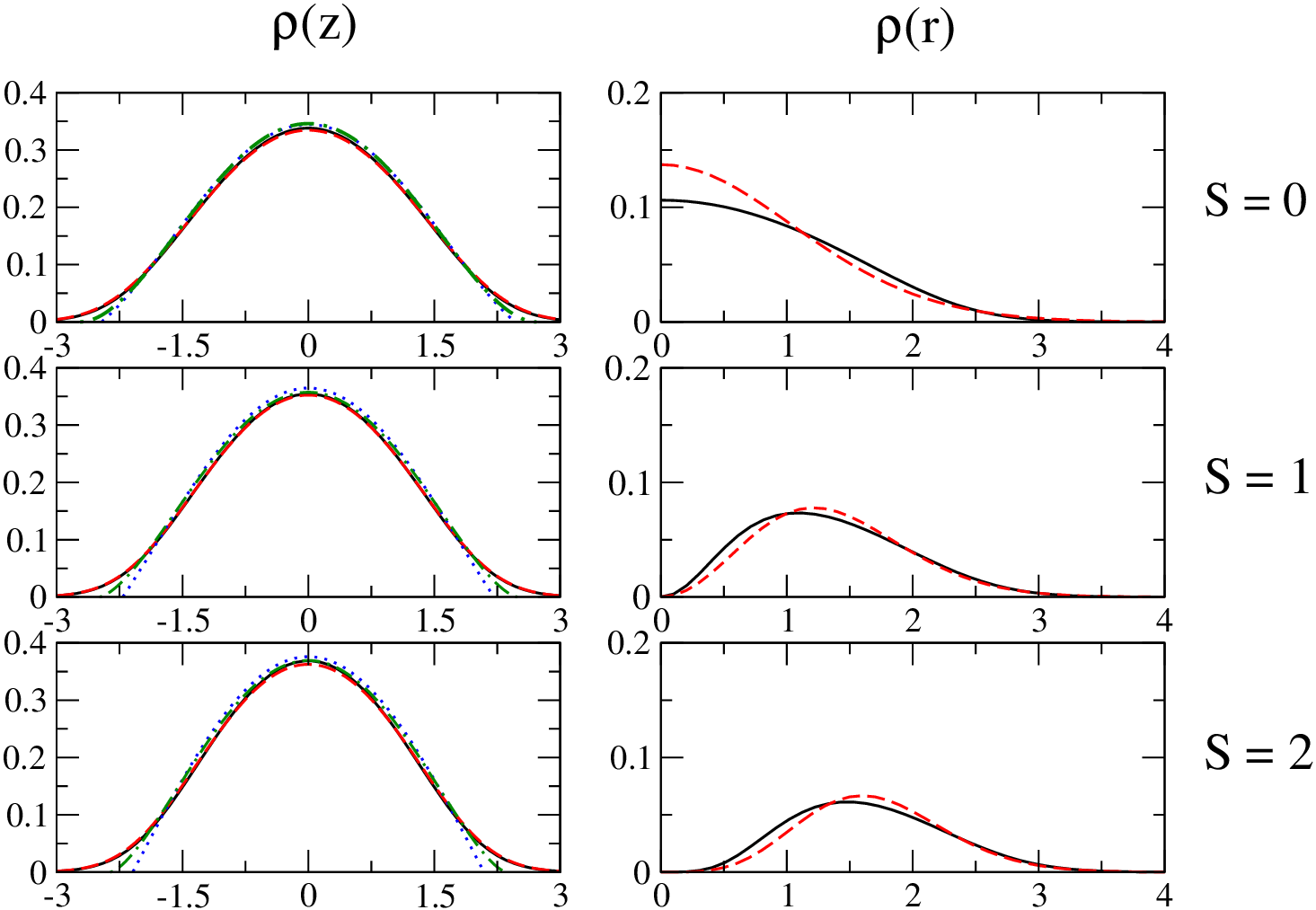}}
\caption{(color online). 
Axial $\protect\rho (z)$ and radial $\protect\rho (r)$ probability
densities of the state with vorticity $S$ in the waveguide, with strength of
self-repulsion $g=20$, and the harmonic potential applied also in the axial
direction, $V(z)=z^{2}/2$. Solid lines: numerical solution of the stationary
3D Gross-Pitaevskii equation (GPE); dashed lines: numerical solution of the
effective one-dimensional NPSE, Eq. (\protect\ref{npse-st}); dotted lines:
obtained the analytical Thomas-Fermi (TF) approximation obtained from the
NPSE, Eq. (\protect\ref{figata}). Dot-dashed lines: the TF approximation for
the 3D GPE.}
\label{fig-g20}
\end{figure}

To test the accuracy provided by the NPSE in this case, in Fig. \ref{fig-g20}
we plot the axial and radial probability densities of the repulsive BEC with
vorticity $S$, defined, respectively, as 
\[
\rho (z)=\int_{0}^{+\infty }|\psi (r,z)|^{2}2\pi r\ dr,~\rho
(r)=\int_{-\infty }^{+\infty }|\psi (r,z)|^{2}\ dz,
\]%
and obtained from the numerical solution 
of the stationary version of Eq. (\ref%
{3DGPE-vortex}). They are compared with predictions of the stationary NPSE (%
\ref{npse-st}) , as well as with analytical result (\ref{figata}) produced
by the TF approximation. We choose $g=20$, and take $V(z)=z^{2}/2$. This
choice of the axial potential implies that the full potential is isotropic.
Thus, we are testing the NPSE in a geometry far from the cigar-shaped one
where it is expected to be very accurate, as it was verified 
in the case $S=0$. 

Figure \ref{fig-g20} shows that, for all $S$, the agreement between the 3D
GPE and the NPSE (solid and dashed lines, respectively) is extremely good in
the axial direction. There are some differences in the radial probability
density, $\rho (r)$. Remarkably, the TF version of the NPSE, i.e., Eq. (\ref%
{figata}) (which corresponds to the dotted-dashed line in the figure) also
provides for very good accuracy of the axial probability-density
distribution in the core of the confined space.

The comparison between the full GPE and the effective NPSE is further
presented in Table 1, which displays values of the ground-state energy of
the repulsive BEC with vorticity $S$, as obtained from the three-dimensional
GPE ($E_{GPE}$), and from the NPSE ($E_{NPSE}$). The former energy is
calculated as 
\begin{equation}
E=\int {\frac{1}{2}}|\nabla \Phi |^{2}+{\frac{1}{2}}(x^{2}+y^{2})|\Phi
|^{2}+V(z)|\Phi |^{2}+\pi g|\Phi |^{4}\ d^{3}\mathbf{r}\;,  \label{energy}
\end{equation}%
where $\Phi (\mathbf{r})$ is the ground-state stationary solution of the GPE
in the 3D setting. The energy $E_{NPSE}$ is also obtained from Eq. (\ref%
{energy}), but with $\Phi (\mathbf{r})$ corresponding to the ground-state
stationary solution of the NPSE. The third column in the table shows that
the relative error $\Delta E_{NPSE}/E_{GPE}$ is quite small, 
and moreover that decreases with the increase of $S$.

\vskip 0.3cm

\begin{center}
\begin{tabular}{|c|c|c|c|}
\hline\hline
~$S$~ & ~$E_{GPE}$~ & ~$\Delta E_{NPSE}/E_{GPE}$~ & $\Delta E_{TF}
/E_{GPE}$ \\ \hline
0 & 3.0729 & 0.0194 & 0.1855 \\ 
1 & 3.6704 & 0.0071 & 0.2558 \\ 
2 & 4.4878 & 0.0070 & 0.2295 \\ \hline\hline
\end{tabular}
\end{center}

\noindent Table 1. {\small Repulsive Bose-Einstein condensate 
with nonlinearity strength }$g=20${\small \ and
vorticity $S$, in axial trapping potential $V(z)=z^{2}/2$. 
$E_{GPE}$ represents the ground-state energy as found 
from the full 3D GPE. 
$\Delta E_{NPSE}/E_{GPE}$ and $\Delta E_{TF}/E_{GPE}$ are the 
relative differences of $E_{GPE}$ with respect to the ground state 
energy of NPSE and TFGPE wave functions.}

\vskip0.3cm

In addition, we have also calculated the size of 
the term $(S+1)\sigma ^{\prime
2}/\sigma ^{2}$ in the effective Lagrangian (\ref{effective}), that was
neglected in the derivation of the NPSE. The result is that the neglected
term is very small indeed, being always less than $0.5\%$ of the energy
produced by the NPSE.

For the sake of completeness, we have also calculated the axial probability
density obtained by the application of the TF approximation to the full 3D
GPE (TFGPE) and subsequent integration over $x$ and $y$. It is given by 
\[
\rho (z)={\frac{1}{2g}}\mu (z)\sqrt{\mu (z)^{2}-S^{2}}-{\frac{S^{2}}{2g}}\ln 
\sqrt{\frac{\mu (z)+\sqrt{\mu (z)^{2}-S^{2}}}{\mu (z)-\sqrt{\mu (z)^{2}-S^{2}%
}}}\;,
\]%
where $\mu (z)\equiv \mu -V(z)$. The left panels of Fig. \ref{fig-g20} show
that the results provided by this TFGPE approximation are
slightly better than those obtained by the same approximation 
applied to the NPSE,
but the full solution of NPSE is better than the TFGPE approximation, 
as shown in the last column of Table 1 and slso in 
the left panels of Fig. 1, where the NPSE curves practically 
coincide with their counterparts produced by the numerical 
solutions of the GPE in 3D. 

\section{Attractive nonlinearity}

Proceeding to the attractive inter-atomic interactions, i.e., negative $g$
(and $g_{S}$), it is relevant to recall that, for $V(z)=0$ and $S=0$, a
family of bright-soliton solutions to Eq. (\ref{npse}) was constructed in
Refs. \cite{sala1,sala2}. In the case of $S\neq 0$, one is dealing with  
\textit{bright vortex solitons}. In a numerical form, they were studied in
Ref. \cite{sadhan}. A Gaussian-based variational approach was 
developed in Ref. \cite{sala-vortex}, and, in a brief form [which resulted
in an effective potential for the vortex soliton in the case of
inhomogeneous transverse confinement, $\omega _{\perp }=\omega _{\perp }(z)$%
] also in Ref. \cite{boris}. The stability of the vortex solitons with both
large and small aspect ratios (i.e., strongly elongated ones, as we consider
here, or pancake-shaped solitons, tightly confined in the axial direction)
was studied by means of accurate numerical methods, based on the
linearization of the GPE with respect to small perturbations, in Refs. \cite%
{DumDum}.

Here we use the NPSE in the form of Eq. (\ref{npse-st}) and compare its
results with those from the full 3D GPE. To simplify the notation, we set 
\[
\gamma _{S}\equiv -|g_{S}|=-|g|\ {\frac{(2S)!}{2^{2S}(S!)^{2}(S+1)},} 
\]
\[
\mu _{S}\equiv {\frac{\mu }{S+1}}\;. 
\]%
By imposing vanishing boundary conditions at infinity, $\phi (z)\rightarrow 0
$ as $z\rightarrow \pm \infty $, we find soliton solutions in an implicit
analytical form, 
\begin{eqnarray}
z\sqrt{2(1+S)} &=&\frac{1}{\sqrt{1-\mu _{S}}}\mathrm{Arctanh} \left( \sqrt{%
\frac{\sqrt{1-\gamma _{S}\phi ^{2}}-\mu _{S}}{1-\mu _{S}}}\right)  \nonumber
\\
&&-\frac{1}{\sqrt{1+\mu _{S}}}\mathrm{Arctan}\left( \sqrt{\frac{\sqrt{%
1-\gamma _{S}\phi ^{2}}-\mu _{S}}{1+\mu _{S}}}\right) \;,  \label{alb}
\end{eqnarray}%
\begin{equation}
\gamma _{S}={\frac{2\sqrt{2}}{3}}(2\mu _{S}+1)\sqrt{1-\mu _{S}}\;.
\label{gamma}
\end{equation}%
The soliton family is then characterized by the dependence of $\gamma _{S}$
on $\mu _{S}$. Eq. (\ref{gamma}), implies that solitons do not exist if the
nonlinearity strength exceeds the critical value, $\gamma _{S}^{\mathrm{(cr)}%
}=4/3$, corresponding to $\mu_{S}=1/2$. For stronger nonlinearities the NPSE
(\ref{npse}) predicts a \textit{longitudinal collapse}, that is a quantum
tunneling to a high-density state \cite{leggett} where atoms quickly
evaporate due to three-body recombination \cite{Cornish}. In terms of the
physical parameters, $a<0$, $a_{\perp }$, and $N$ (number of atoms), we
conclude that the collapse of the bright vortex soliton takes place in
region 
\begin{equation}
{\frac{N|a|}{a_{\bot }}}>{\frac{2}{3}}{\frac{2^{2S}(S!)^{2}(S+1)}{(2S)!}}.
\label{collapse}
\end{equation}%
In the opposite limit of weak nonlinearity, $\gamma _{S}\rightarrow 0$,
exact soliton solution (\ref{alb}) takes the ordinary form, 
\[
\phi (z)=\sqrt{\frac{\gamma _{S}}{4}}\mathrm{sech} \left( {\frac{1}{2}}%
\gamma_{S}\sqrt{S+1}z\right) \;. 
\]

As a final remark, we note that the inversion of Eq. (\ref{gamma}) provides
two branches of solutions for $\mu_{S}(\gamma_{S})$. In numerical
simulations of Eq. (\ref{npse}), only the one satisfying the condition $%
d\mu_{S}/d\gamma_{S}<0$ turns out to be stable, in precise agreement with
the prediction of the \textit{Vakhitov-Kolokolov} stability criterion \cite%
{VK}.

\begin{figure}[tbp]
{\includegraphics[height=2.3in,clip]{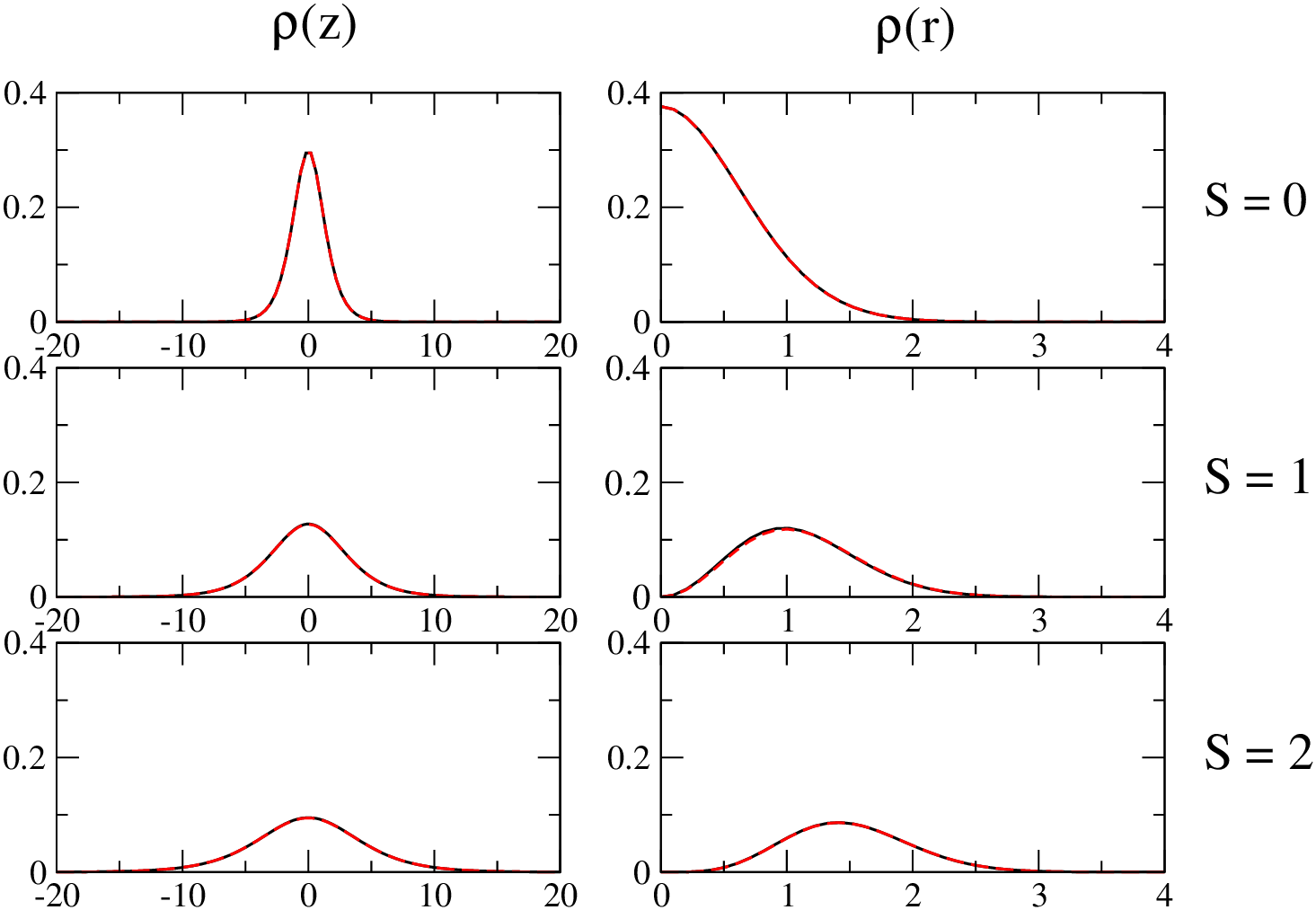}}
\caption{(color online). 
Axial $\protect\rho (z)$ and radial $\protect\rho (r)$ probability
densities in the soliton state with vorticity $S$, in the case of the
self-attraction with strength $g=-1$ and $V(z)=0$ (no axial potential).
Solid lines: obtained from the full 3D GPE. Dashed lines: produced by the
NPSE.}
\label{fig-g1-open}
\end{figure}

In Fig. \ref{fig-g1-open} we plot the axial probability density $\rho (z)$
of the soliton with vorticity $S$, with no axial external potential ($V(z)=0$%
), and for $g=-1$. The figure shows that the radial probability density $%
\rho (r)$ produced by the NPSE is virtually indistinguishable from that
generated by the 3D GPE. This can be understood since the system with $g=-1$
becomes effectively quasi-one-dimensional. The figure also shows that, for
all $S$, the agreement between the 3D GPE and the NPSE (solid and dashed
lines) is excellent also in the axial direction. Thus, we conclude that the
NPSE predicts the bright \emph{vortex solitons} with a very high accuracy.

\section{Vortex bright solitons in a ring}

Repulsive BECs in a ring-shaped (toroidal) trap are now available to the
experiment \cite{r5}. Self-attractive BEC in a ring has not been
experimentally created yet, but it may also be a promising setting,
interesting from a physical point of view, since a quantum phase transition
from a uniform state to a bright soliton has been predicted in Refs. \cite%
{r6}. This prediction is based on the mean-field and beyond-mean-field
numerical analysis for the 1D Bose gas with contact interactions and
periodic boundary conditions. Recently, self-attractive BECs in a 3D
ring-shaped trap were considered, taking into account the transverse
structure of the trapped condensate. In this way, new features in the
model's phase diagram have been predicted, both at zero \cite{sala-ring} and
finite \cite{sala-ring2} temperatures.

The analysis presented in Refs. \cite{sala-ring,sala-ring2} can be extended
to the bright solitons with intrinsic vorticity by using the above-mentioned
scaling that makes Eq. (\ref{npse-st}) equivalent to its counterpart with $%
S=0$. As shown in Ref. \cite{sala-ring}, one can model the BEC in the tight
3D ring by using Eq. (\ref{npse-st}) with natural periodic boundary
conditions (b.c.), $\phi (z+L)=\phi (z)$, with $-L/2<z<L/2$, where $L=2\pi R$
is the length of the ring with radius $R$ (here, we do not assume any global
vorticity imposed along the closed ring).

Equation (\ref{npse-st}) subject to the periodic b.c. always admits the
axially uniform solution, $\phi =1/\sqrt{L}$. Further, in Ref. \cite%
{sala-ring} it was shown that this uniform state, with $S=0$, is
energetically and dynamically stable in the attractive BEC only for a finite
range of parameters. In the case of $S\neq 0$, the stability condition is
modified, by the above-mentioned scaling transformation, to 
\[
{\frac{\pi ^{2}}{\gamma _{S}L\sqrt{S+1}}}\left( 1-\frac{\gamma_{S}}{L\sqrt{%
S+1}}\right) ^{3/2}\geq \left( 1-\frac{3\gamma _{S}}{4L\sqrt{S+1}}\right) , 
\]%
which for large $L$ reduces to 
\begin{equation}
{\frac{\pi ^{2}}{\gamma _{S}L\sqrt{S+1}}}\geq 1.  \label{largeL}
\end{equation}%
In terms of physical parameters, condition (\ref{largeL}) takes the
following form, cf. condition (\ref{collapse}) for the onset of the collapse
in the rectilinear trap: 
\begin{equation}
{\frac{N|a|}{a_{\bot }}}\leq {\frac{\pi ^{2}a_{\bot }}{2L}}{\frac{
2^{2S}(S!)^{2}\sqrt{S+1}}{(2S)!}}  \label{transition}
\end{equation}
Note that in this inequality both the scattering length $a$ and $L$ are
expressed in dimensional units.

\begin{figure}[tbp]
\vskip0.5cm {\includegraphics[height=3.2in,clip]{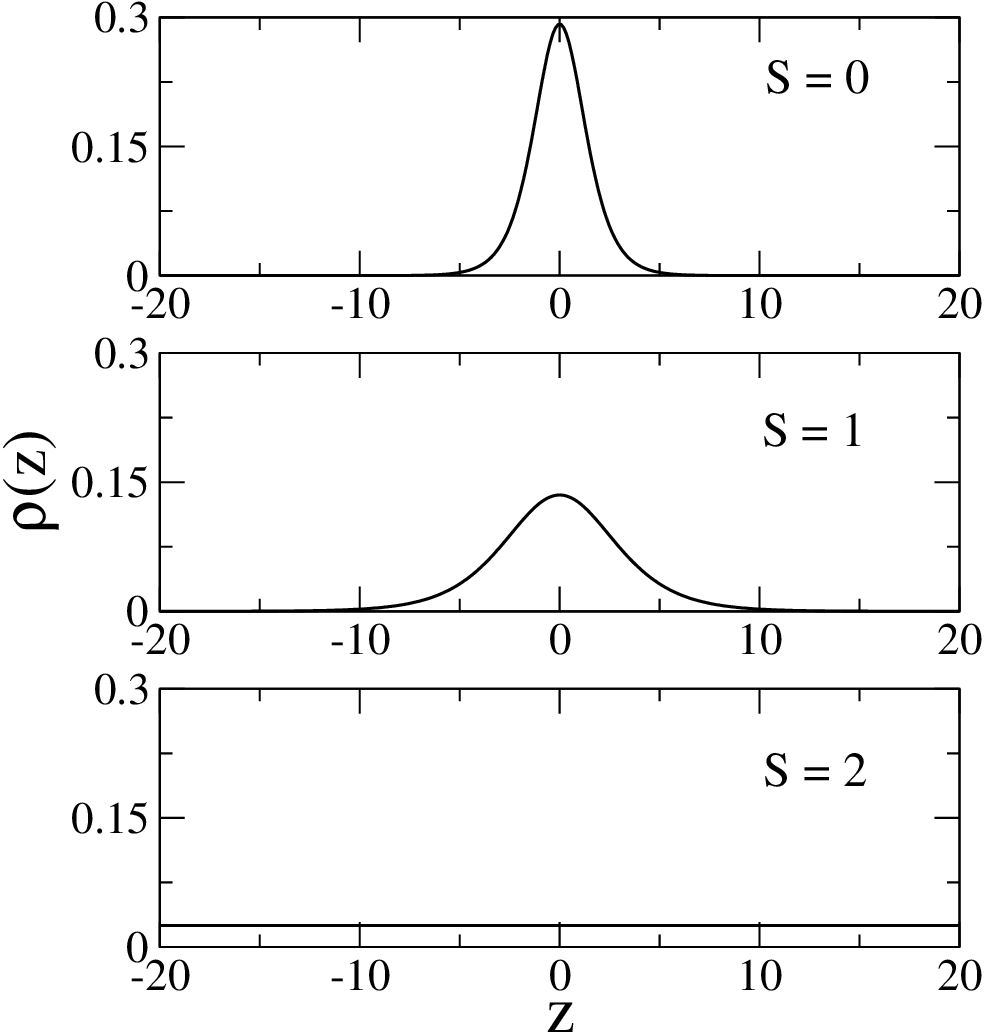}}
\caption{Axial probability density $\protect\rho (z)$ in the state with
transverse vorticity $S$, in the ring-shaped waveguide, i.e., with the
periodic boundary conditions in $z$. Self-attraction strength is $g=-1$, and
the length of the ring is $40$.The results are obtained from the NPSE.}
\label{fig-g1}
\end{figure}

In Fig. \ref{fig-g1} we plot several profiles of the axial probability
density $\rho (z)$, as obtained from the NPSE, Eq. (\ref{npse}), with
vorticity $S$ and $g=-1$ [with no external axial potential: $V(z)=0$].
Periodic boundary conditions are imposed, with period (length of the ring) $%
L=40$. We have checked that, as well as in Fig. \ref{fig-g1-open}, the
agreement between results produced by the 3D GPE and NPSE is very good for
all $S$ (we display only the profiles generated 
by the NPSE since they are
indistinguishable in the figure from their counterparts obtained from the 3D
equation). Note that, contrary to the case shown in Fig. \ref{fig-g1-open},
in the present case the axial probability density is uniform for $S=2$, due
to the delocalization transition, the threshold for which is very accurately
predicted by Eq. (\ref{transition}).

\section{Conclusions}

In this work, we have presented a systematic analysis of self-repulsive
and self-attractive matter-wave patterns with the intrinsic vorticity which
fill a straight or toroidal cigar-shaped trap. The 1D NPSE 
was derived from the underlying 3D GPE, as a modification of 
the previously derived 1D 
equation for the states without vorticity. In the cases of the
self-repulsion and self-attraction, the modified NPSE predicts,
respectively, the axial density profiles for the TF states
and bright solitons, both in an analytical form. The comparison of these
predictions with direct numerical solutions of the underlying 3D GPE for the
vortex states demonstrates a very high accuracy of the description provided
by the NPSE, which includes the prediction of the stability of the bright
solitons, as well as the threshold for the onset of collapse. For the
toroidal configuration of the trap, a threshold for the transition from the
axially uniform vortex state to a quasi-soliton symmetry-breaking pattern
was found too.

This work may be naturally extended in several directions. In particular, a
challenging problem is collision of solitons with opposite signs of the
intrinsic vorticity, $+S$ and $-S$. This setting cannot be described by the
NPSE (which assumes a single value of $S$ throughout the elongated trap),
hence it should be studied in the framework of the full 3D GPE. Another
interesting generalization may be formation of gap solitons with intrinsic
vorticity, in the case when the nonlinearity is self-repulsive, and the
cigar-shaped trap is combined with a periodic axial potential, like in the
well-known experimental setting used for the creation of ordinary gap
solitons \cite{Markus}.

This work has been partially supported by Fondazione CARIPARO. L.S. has been
partially supported by GNFM-INdAM and thanks Alberto Cetoli for 
useful discussions.

\end{document}